# Ge thin-films with tantalum diffusion-barriers for use in Nb-based superconductor technology

Cameron Kopas, Shengke Zhang, J. Gonzales, Daniel R. Queen, Brian Wagner, R. W. Carpenter, N. Newman

## Abstract

Germanium thin films are an excellent candidate for use as a low-loss dielectric in superconducting microwave resonators, a low-loss inter-layer metal wiring dielectric, and passivation layers in microwave and Josephson junction devices. In Ge/Nb structures deposited at 400 °C, we observe intermixing over as much as 20 nm. The addition of a 10 nm Ta diffusion barrier layer reduces the superconductor/dielectric intermixing to less than 5 nm and enhances the structural properties of deposited *a*-Ge layers based on Raman spectroscopy. Additionally, superconducting microwave resonators fabricated at room-temperature on crystalline Ge substrates with a Ta barrier layer show marked improvement in total and power-dependent two-level system microwave losses.

## Introduction

Germanium has many properties that make it appealing for use as an insulating dielectric for low-temperature superconductor devices. Germanium can be produced with impurity concentrations as low as $10^8$ cm$^{-3}$, which is lower than high resistivity float zone silicon where the impurity concentration is $\sim 10^{15}$ cm$^{-3}$, with the majority from Si-O complexes [1]. Additionally, high-quality Ge can be deposited directly on Si via sputtering or evaporation with low surface roughness and low dangling bond density. Germanium films grown at T~30 °C are amorphous while high-temperature growth yields polycrystalline (T>400 °C) and epitaxial films (T>450 °C) [2]. While undoped Ge has immeasurably low conductivity below 50 K, it has sufficient conductivity at room temperature to protect against electrostatic discharge during device fabrication and storage.

Superconducting qubits operating at low temperatures (~40 mK) and near single-photon power are sensitive to losses due to two-level systems (TLS) found at the metal-air, metal-dielectric, and dielectric-air interfaces [3,4,5]. TLS loss limits performance in superconducting microwave devices such as co-planar waveguides (CPW) and Kinetic Inductance Detectors (KIDs), highlighting the importance of high-quality interfaces. Identifying the interface defects that give rise to these TLS is the first step in a holistic approach to improving quantum device performance by engineering the interface states.

In an earlier study [6] we demonstrated very low TLS losses (tan $\delta_0$=1.3×10$^{-6}$) in Nb thin-film resonators on room-temperature deposited Ge dielectric films. Our more recent investigations of Nb/Ge indicate that some properties may be improved by depositing Ge at temperatures that promote crystallization (~400 °C for Ge on Si [2]), but only when the effects of interfacial intermixing could be reduced or eliminated. Some previous works use alloy superconductors for CPW (e.g., TiN and NbTiN on Si or $Al_2O_3$) [7,8]; however, we are not aware of studies involving diffusion barriers for Nb-layers on Si or Ge.

Tantalum is a likely diffusion barrier layer. It is a refractory metal with a similar body-centered cubic (BCC) crystal structure and nearly-identical lattice constant to niobium ($a_{Nb}$=330.04 pm, $a_{Ta}$=330.13 pm). While the Ta superconducting critical temperature ($T_c$ = 4.4 K) is lower than that for Nb ($T_c$= 9.2 K), a 10 nm barrier layer (thinner than the Ta coherence length $\xi_{Ta}$ ~95 nm, $\xi_{Nb}$=38 nm) will not significantly degrade the $T_c$ of a Nb/Ta bilayer. In the clean limit where interface reflection is zero, we would expect the $T_c$ of a Nb (200 nm)/Ta (10 nm) bilayer depressed at most from 9.2 K to 8.6 K [9]. Furthermore, microwave device studies of Nb and Ta co-planar microwave resonators indicate that similarly prepared devices do not differ in phase noise at the same $T/T_c$ [10].

The interface between sputtered niobium and crystalline silicon is known to have 10-20 nm of intermixing with additional thermal processing resulting in complete silicidation [11,12]. In this study, similar levels of intermixing occur in the Nb/Ge system without a barrier layer. We show that a Ta barrier can be used to reduce interdiffusion at the niobium-germanium interface. We also characterize and then compare the structural properties of amorphous and polycrystalline Ge layers deposited on Nb and Ta layers.

**Experimental Procedures**

Thin-film deposition (Nb, Ta, and Ge) was performed in a cryopumped (Model CTI-8, CTI, Austin, TX) unbaked UHV vacuum system with a base pressure of less than $2\times10^{-9}$ Torr. Float-zone Si (100) substrates were cleaned ultrasonically for 10 minutes each in semiconductor-grade acetone and then ethanol followed by a 5 minute 2 vol % HF aqueous solution etch. Undoped zone-refined Ge (100) wafers with stated resistivity higher than 40 Ω-cm were used as both substrates and the source of evaporation material.

Niobium and tantalum were deposited via magnetron sputtering from 5 cm diameter, 99.95% pure targets. D.C. magnetron sputter deposition was initiated after evacuating to the base pressure, then backfilling to 4 mTorr of 99.9999% Ar. Nb and Ta were deposited with a sputtering power of 215 W, corresponding to a deposition rate of ~6 Å/second for Nb and a thickness of ~200 nm, and ~4 Å/second for Ta and a thickness of 10-15 nm. Germanium was evaporated from a Knudsen effusion cell. After the metal sputter deposition, germanium film deposition was started after evacuating the chamber to below $2\times10^{-9}$ Torr. The germanium evaporation source was heated to 1400 °C for a deposition rate of ~2.4 Å/second and a total Ge thickness of ~1 μm.

Rutherford backscattering spectrometry (RBS) was used to measure the layer thickness and monitor for contaminants. Raman spectroscopy was performed using a 532 nm laser operating at 0.75 mW and 0.5 μm spot size. X-ray Diffraction was performed using a high-resolution diffractometer with an X-ray mirror on the incident optic, and a 0.27° parallel plate collimator on the receiving optics (PANalytical X'pert MRD Pro).

Transmission Electron Microscopy (TEM) cross-section specimens were prepared using two different methods. In specimens with a top Ge layer, cross sections were prepared by manual "rod and tube" preparation to avoid removal of the upper Ge layer. Specimens were bonded with epoxy inside a brass support structure, diced into thin cross-sections, mechanically polished down to less than 100 μm thickness, thinned to a few μm thick in the center using a dimple grinder, and polished to electron transparency using a Precision Ion Polishing System (PIPS) mill. These specimens were characterized using an aberration-corrected Jeol ARM-200F in STEM configuration with in-situ Energy Dispersive X-ray Spectroscopy (EDS), and an AppFive precession electron diffraction system. Specimens with a Nb top layer were fabricated using the same structures used for the microwave resonator measurements. These cross-sections were prepared by focused ion beam milling using an FEI Helios 660 Dual-Beam FIB/SEM. These specimens were measured using an aberration-corrected Hitachi 2700 STEM.

Superconducting co-planar microwave waveguide resonator (SCPW) fabrication used standard photolithography techniques and $CF_4$ reactive ion etching, creating a 5.8 GHz stripline resonator with 4 μm trace width, and 2 μm gap width. The resonator's electric field was concentrated in the deposited dielectric layer by using a narrow trace width and keeping the over-etch to less than 20 nm. Wider geometries are preferred for quantum devices so that more electric field energy is stored in the bulk substrate than the interfaces to maximize total quality. We measure transmission measurements ($S_{21}$) at ~40 mK as a function of applied power. The resonance is fit using the diameter correction method to extract $Q_i$, the internal quality factor of the resonator and the loss tangent ($\tan \delta_i = Q_i^{-1}$) [13]. The power-dependent loss tangents are fit to the two-level system model [14].

## Results and Discussion

The bright-field STEM micrograph in Figure 1-a shows the interface between Ge evaporated at 400 °C and the Nb film underneath. Moiré fringes in the Ge layer represent multiple grains through the thickness of the sample region and not structural information about the interface. Lattice planes are visible extending across the interface near the right side of the image indicating regions with 5-10 nm of intermixing. Precession diffraction (Figure 1-b) finds 150 nm thick Ge grains that are on average ~50 nm in width. XRD and TEM analysis both find that the deposited Ge layer is polycrystalline with Ge (111), (220), and (311) planes oriented perpendicular to the Nb surface. The observation of significant intermixing in the high-magnification STEM image prompted the investigation into diffusion barrier layers.

We quantify the extent of intermixing in structures with high-temperature deposited Ge using energy-dispersive x-ray spectroscopy (EDS) line profiles (Figure 2). Ge evaporated at 400 °C directly onto Nb exhibits as much as 20 nm of intermixing between the Nb and Ge layers. For a Ge layer deposited at 400 °C onto Ta-buffered Nb, we observe less than 5 nm of intermixing, significantly less than the intermixing across the Ge/Nb interface. Note that the EDS probe size is between 2–5 nm because of beam broadening through the specimen thickness. The tall columnar grain structure of the niobium and tantalum is visible in the STEM images; the contrast differences from the bottom to the top of the Nb grains are due to variable specimen thickness. The Ta has a uniform thickness and conformally coats the Nb, matching the grain sizes and orientation of the underlying Nb. These results show that Ta is an effective diffusion barrier in niobium-germanium stacks, even at temperatures as high as 400 °C.

Figure 3 shows cross-section STEM micrographs of Nb sputter-deposited at room-temperature onto Ge substrates with and without a Ta barrier layer. EDS line profiles measured across the interface find ~8 nm of intermixing between Nb and Ge. In the structure with a Ta barrier layer, the Ge composition drops from 100% to 10% within ~6 nm, then 10% to 0% in ~10 nm. While the total extent of intermixing is ~16 nm, there is no evidence of metals (Nb or Ta) diffusing into the Ge layer past 6 nm, and a trace amount of Ge diffused into the superconducting layer is not expected to significantly affect the superconducting properties of the layers.

We infer the relative structural quality of amorphous (room-temperature grown) and polycrystalline (400 °C grown) Ge films deposited onto superconductor layers by comparing the Raman peak shift and full-width half maximum (FWHM) of the Ge transverse optic-like (TO-like) mode (shown in Figure 4) to a reference Raman spectra of undoped single-crystal Ge (100). Ge deposited on a Ta diffusion barrier has a narrower peak, as quantified by the full width at half maximum (FWHM), and peak maximum wavenumber closer to the reference peak value in both amorphous and polycrystalline Ge. This indicates that there is less strain in the Ge layers deposited on Ta buffer layers than that when deposited directly on Nb [15]. The narrower peak is a result of a longer phonon lifetime, as would be expected for higher structural quality films. The lower-intensity broad Raman peaks observed for the amorphous films are similar to those found in electrolytically deposited *a*-Ge reported in the literature [16].

The temperature-dependent resistivity characteristics of a parallel plate structure Nb/Ta/Ge(400 °C)/Ta/Nb/Si substrate, as shown in Figure 4, are dominated by the resistance of the top Nb layer. We find a $T_c$ of 9.2 K and a Residual Resistivity Ratio (RRR) of 6.6 that is typical of high-purity and high-structural quality Nb films deposited at room-temperature on Si substrates. Using the Sommerfeld model, we infer a mean free path at 10 K of 7 nm and no significant depression in $T_c$ or increase in kinetic inductance from the Ta layer.

Low-temperature (40 mK) microwave measurements of Nb (200 nm)/Ta (10 nm)/undoped single-crystal Ge resonators exhibit single photon loss tan $\delta_i$ = 4.92×10$^{-5}$. This loss value is lower than the corresponding loss for Nb directly on crystalline Ge, with tan $\delta_i$ of 7.53×10$^{-5}$. These measurements indicate that a barrier layer can improve the performance of microwave resonators.

## Conclusions

We have demonstrated that a Ta diffusion barrier between Nb and Ge decreases intermixing, improves microwave loss properties, and improves the structural quality of deposited Ge films. The Ta diffusion barrier reduces interface intermixing between Ge and Nb from ~20 nm to less than 5 nm in structures fabricated at temperatures at 400 °C. Low-temperature (40 mK) and low power measurements of SCPW resonators deposited on crystalline Ge wafers find improved TLS and total microwave loss properties, attributed to the addition of Ta diffusion barrier layer (i.e., tan $\delta_i$ = 4.92×10$^{-5}$ with Ta-diffusion barrier, and without a diffusion barrier tan $\delta_i$ = 7.53×10$^{-5}$. The combination of these properties enables the production of high Q, low TLS loss co-planar, microstrip and stripline resonators, as well as low-loss device isolation, inter-wiring dielectrics, and surface passivation layers in passive microwave and Josephson junction circuit fabrication.

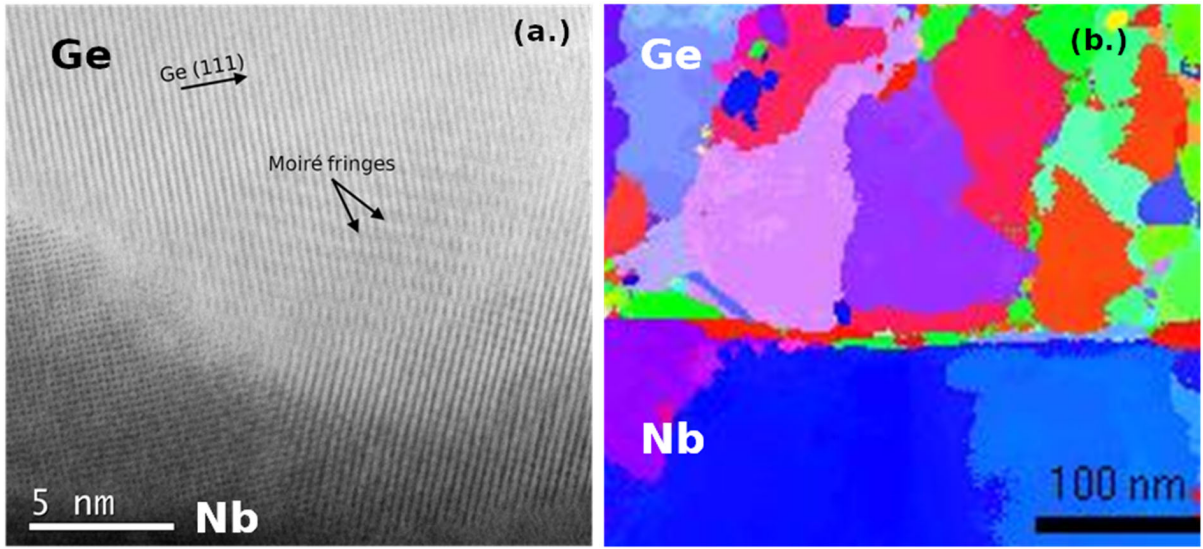

Figure 1 (a.) High-resolution bright-field STEM image showing the Ge-Nb interface for Ge evaporated onto Nb at 400 °C. (b.) Precession diffraction across the Nb/Ge interface showing a colorized grain contrast of the Ge layer deposited onto Nb at 400° C. `

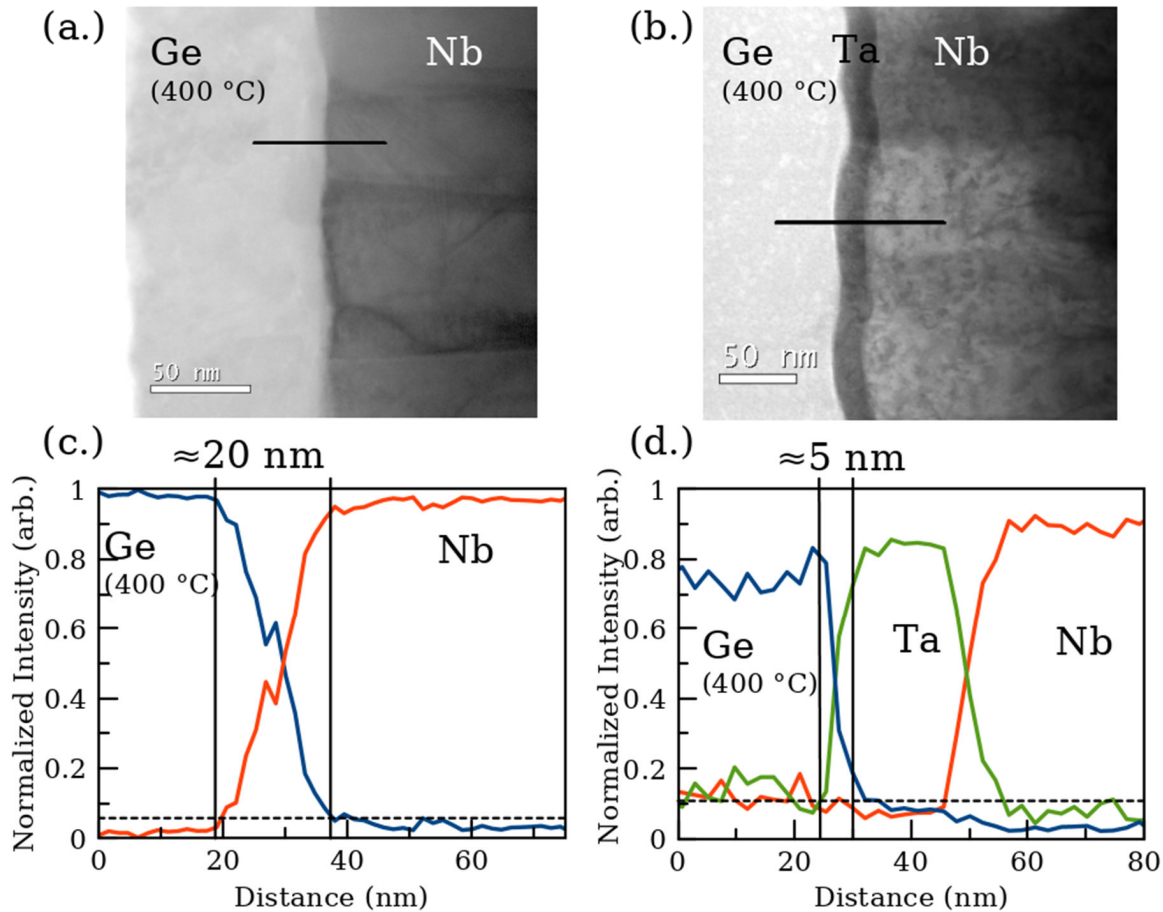

Figure 2: Bright-field STEM cross-section images across the Ge/superconductor interfaces, and corresponding EDS line profiles across these interfaces. The approximate area of the EDS line profile is indicated by a black line in the images. (a.) bright-field STEM image of Ge evaporated onto Nb at 400 °C, (b.) Bright-field STEM image where Ge was evaporated onto Ta on Nb film at 400 °C. (c.) The results of an EDS line profile across the Ge/Nb interface, marked by the line in (a.) and showing ~20 nm of Ge/Nb intermixing. (d.) The results of an EDS line profile across the Ge/Ta/Nb interface, from the region marked by the line in (b.), showing ~5 nm of Ge/Ta intermixing, near the EDS probe resolution of 1.5 nm. The dotted lines correspond approximately to the EDS background.

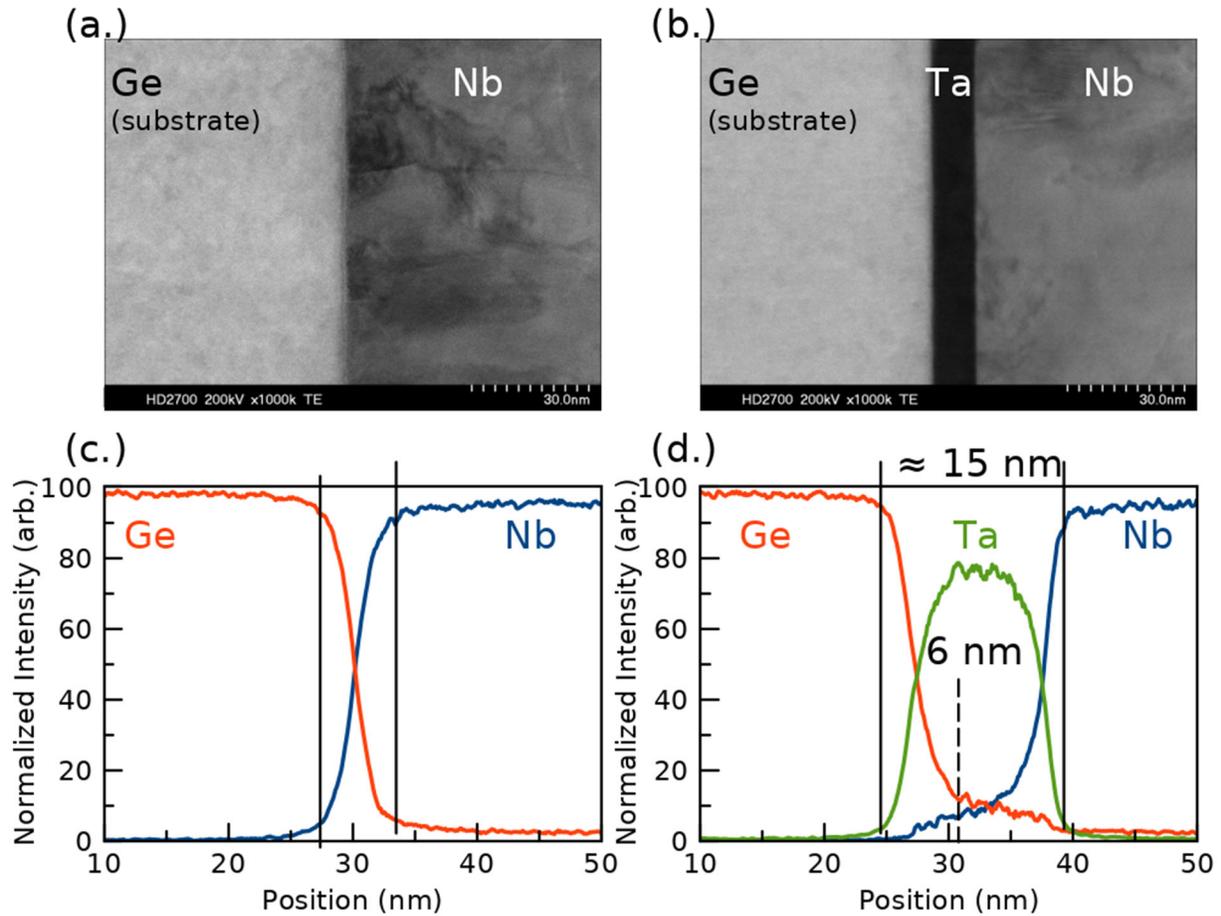

Figure 3: Bright-field STEM cross-section images of superconducting films sputter deposited on crystalline Ge substrates, and corresponding EDS line profiles across these interfaces. (a.) Bright-field STEM image of the Ge/Nb interface. (b.) Bright-field STEM image of the Ge/Nb/Ta interface. (c.) The results of an EDS line profile across the Ge/Nb interface shown in (a.). (d.) The results of an EDS line profile across the Ge/Ta/Nb interface, from the region in (b.)

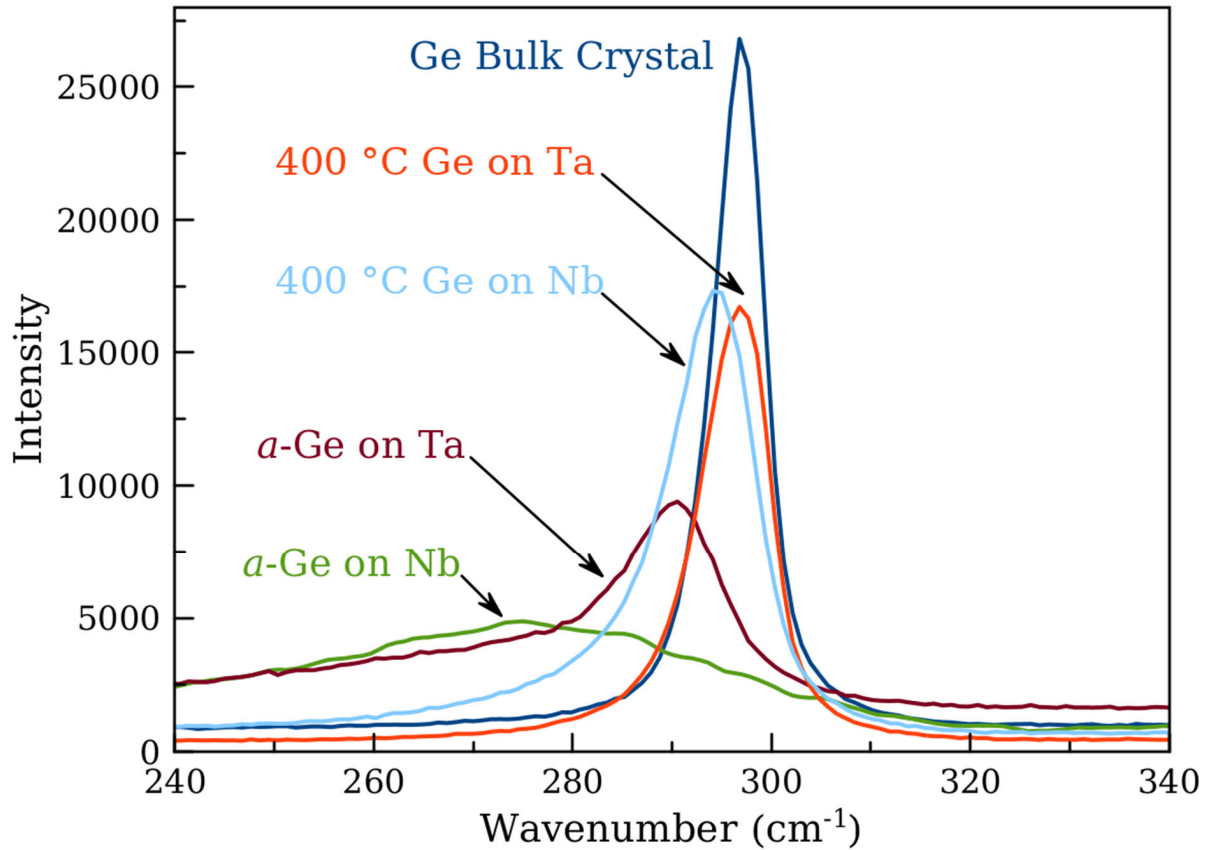

Figure 4: Raman Spectroscopy of Ge TO mode for Ge deposited on superconductor films compared with a crystalline (100) Ge substrate. The Ge on Ta film deposited at 400 C exhibits a peak position closest to that of bulk single-crystal material.

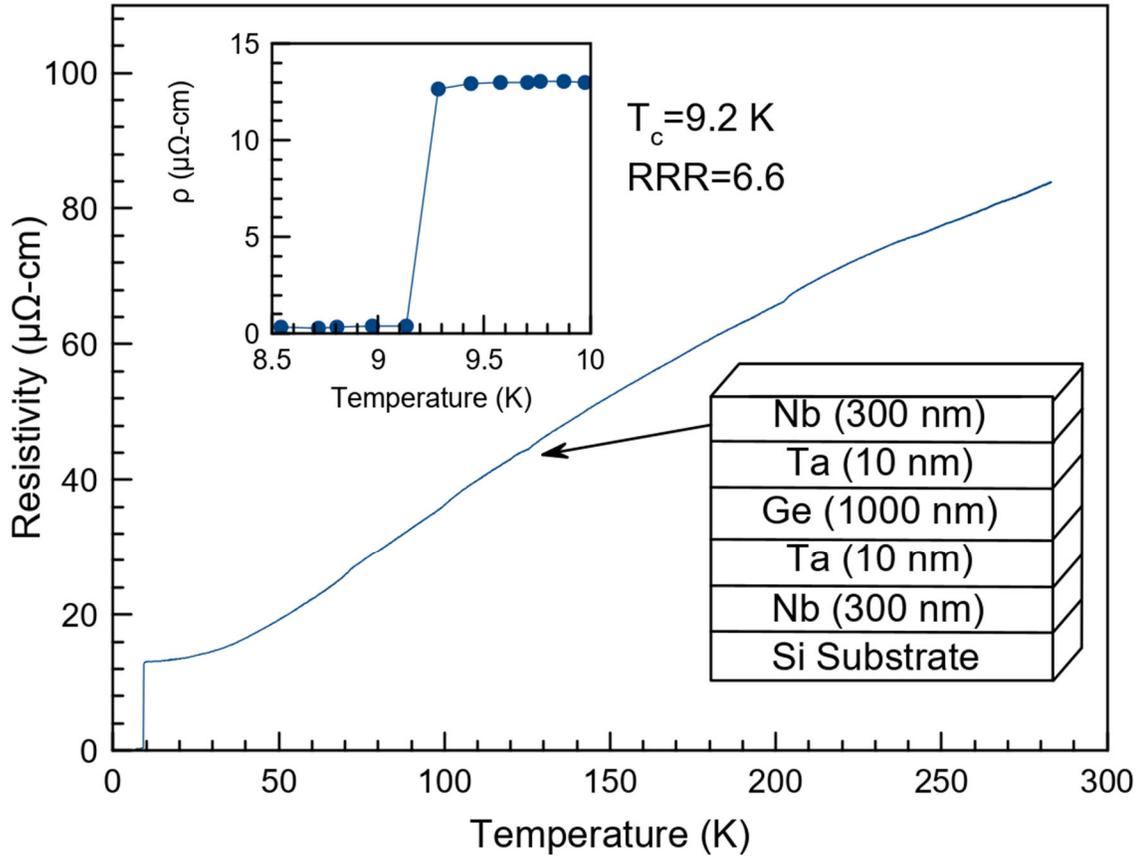

Figure 5: Resistivity vs. temperature measurements measured via 4-point probe on the top surface of a Nb(200 nm)/Ta (10 nm)/Ge (1 μm)/Ta (10 nm)/Nb (200 nm)/Si substrate structure with Ge deposited at 400° C. The superconducting transition temperature, $T_c$, is 9.2 K, and Relative resistivity ratio, RRR, is 6.6.

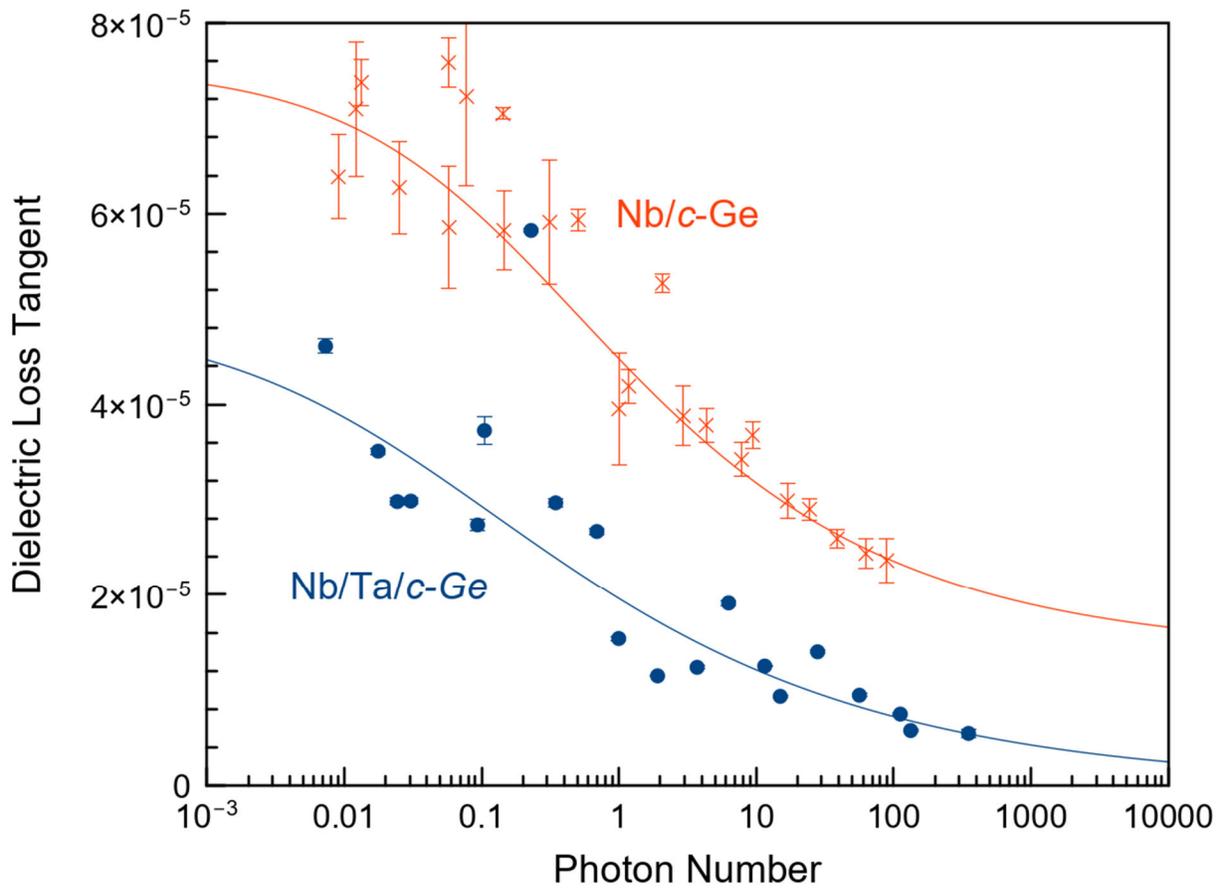

Figure 6: Low-power dielectric loss tangents (tan δ) measured at low temperature and low power for patterned SCPW devices of niobium on undoped single-crystal Ge wafers with and without Ta diffusion barriers. Each resonator has 4 μm trace widths and 2 μm gap widths. The Nb (200 nm)/ Ta (10 nm) structure on has total loss $\tan \delta_i = 4.92 \times 10^{-5}$. The Nb (200 nm) directly on single-crystal Ge (presented earlier in [6] ) has total loss $\tan \delta_i = 7.53 \times 10^{-5}$.